\documentstyle[12pt]{article}

\def\inD{S}
\def\outD{T}
\def\DQ{\mbox{QQC}}

\def\part{\pi}

\begin{document}

\date{}

\newtheorem{theorem}{Theorem}
\newtheorem{lemma}{Lemma}
\newtheorem{definition}{Definition}
\newtheorem{corollary}{Corollary}
\newtheorem{fact}{Fact}
\newtheorem{example}{Example}

\title{On the Quantum Query Complexity of Detecting Triangles
in Graphs} 

\author{
Mario Szegedy\thanks{The research was supported by NSF grant 0105692}\\
Rutgers University}
\maketitle

\thispagestyle{plain}

\begin{abstract}
We show that in the quantum query model the complexity of
detecting a triangle in an undirected graph on $n$ nodes
can be done using $O(n^{1+{3\over 7}}\log^{2}n)$ quantum queries.
The same complexity bound applies for outputting the triangle
if there is any. This improves upon the earlier bound of 
$O(n^{1+{1\over 2}})$.
\end{abstract}

\section{Introduction}

The classical black box complexity of graph properties 
has made its fame through the notoriously hard evasiveness
conjecture of AAnderaa, Karp and Rosenberg (see e.g. \cite{ly})
which states that every non-trivial Boolean function on graphs 
whose value remains invariant 
under the permutation of the nodes has
deterministic query complexity exactly ${ n\choose 2}$, where $n$
is the number of nodes of the input graph.

The question of
quantum query complexity of graph properties was first 
raised in \cite{bcwz}. Here the authors show that 
the quantum model behaves differently from the deterministic
model in the zero error case, although
an $\Omega(n^{2})$ lower bound still holds. 

The bounded error quantum query model is more analogous to the classical
randomized model. About the latter we know much less, in particular 
the general lower bounds are far from the 
conjectured $\Omega(n^{2})$. For a 
long time Peter Hajnal's $\Omega(n^{4/3})$ bound \cite{h} 
was the best, until it was slightly
improved in \cite{ck}. The best general lower bound, $\Omega(n^{2/3})$,
in the bounded error quantum case is due to
Yao and Santha (not yet published).

The quantum query complexity of the Triangle Detection Problem was
first treated in H. Buhrman, C. D\"urr, M. Heiligman, 
P. H$\not${o}yer, F. Magniez, M. Santha and R. de Wolf \cite{bdhh},
where the authors show that in the case of sparse graphs 
the folklore $n^{1.5}$ upper bound can be improved.
Their method breaks down when
the graph has $\theta(n^{2})$ edges. The folklore bound is 
based on searching the space of all
$n^{3}$ triangles with Grover's algorithm \cite{g}. 
Since the above search
idea is quite natural, one might conjecture that the upper bound is 
the best possible.

Below we show however, using combinatorial ideas, that
a quantum query machine can solve the
Triangle Detection Problem in time
$O(n^{1+{3\over 7}}\log^{2}n)$. The lower bound remains $\Omega(n)$.

\section{Notations}

We denote the set $\{1,2,\ldots,n\}$ by $[n]$.
A a simple undirected graph is a set of edges $G\subseteq
\{(a,b)\mid \; a,b\in [n];\; a\neq b\}$ with the understanding that
$(a,b) \stackrel{def}{=} (b,a)$. 
The complete graph on a set $\nu\subseteq [n]$ is denoted by $\nu^{2}$
The neighborhood of a $v\in [n]$
in $G$ is denoted by $\nu_{G}(v)$. Its definition:
\[
\nu_{G}(v) = \{b\mid \; (v,b)\in G\}.
\]

\noindent We denote $|\nu_{G}(v)|$ by $\deg_{G} v$.
For sets $A,B\subseteq [n]$ let

\begin{eqnarray*}
G(A,B) & = & \{(a,b)\mid \; a\in A;\;b\in B;\;(a,b)\in G\}.
\end{eqnarray*}

The following function will play a major role in our proof:
We denote the number of paths of length two from $a\in [n]$
to $b\in [n]$ in $G$ with $t(G,a,b)$. In formula:

\[
t(G,a,b) = |\{x\mid \; (a,x)\in G;\; (b,x)\in G \}|.
\]

\noindent 
$t(G)$ is the number of triangles in $G$.
We define an operation on graphs. For a graph $G\subseteq [n]^{2}$
define

\[
G^{\langle t \rangle} = \{ (a,b)\in [n]^{2}\mid \; t(G,a,b) \le t \}.
\]

\section{The Triangle Detection Problem (TDP)}

\noindent{\bf Input:} An undirected graph $G$ with vertex set $[n]$,
given by its incidence matrix, $G(a,b)$. ($G(a,b)$ is a 
symmetric zero-one valued matrix with all zeros in the diagonal.)

\medskip

\noindent{\bf Output:} ``No,'' if $G$ is triangle free.
Else, a set $\{a,b,c\}$ such that 
$(a,b), (b,c), (c,a) \in G$.

\medskip

Any classical machine that solves the Triangle Detection Problem (TDP) has to 
query all edges of the graph, and even a classical 
randomized machine has to query a
constant fraction of all edges. In contrast, it has been known that 
a quantum query machine can successfully output a triangle of $G$, if exists,
making only $O(n^{1.5})$ quantum queries to $G$,
with high probability. In this section we give an algorithm 
that improves on the exponent $1.5$.

The basis states of a quantum query machine for the TDP are of the form 
$|a,b \rangle |c\rangle$, where $1\le a\le b\le n$ and $c\in S$ is
specific to the the machine.
The starting state is $|1,1 \rangle |c_{0}\rangle$ for some 
distinguished $c_{0}\in S$.
A query step is a unitary operator $O=O_{G}$ acting 
on the basis elements as:
\[
O_{G}|a,b \rangle | c \rangle = (-1)^{G(a,b)} |a,b \rangle | c \rangle. 
\]

The computation is a sequence $U_{0}O_{G}\ldots U_{t-1}O_{G}U_{t}$
of query steps and 
non-query steps. Each non-query step is an arbitrary 
unitary operator on the state space with the only restriction that
it does not depend on $G$.
The output is read from a measurement of the final state,
and needs to be correct with probability at least $1-\epsilon$.
We will describe our algorithm as a one that runs on a 
classical machine that calls quantum subroutines, but makes
classical queries as well. Although this type of the machine seems more
general than the one we explained above, the two models 
can be shown to have the same power.

The quantum query complexity of the
Triangle Detection Problem as well as of many of its kins with small 
one-sided certificate size are notoriously hard to analyze, because
one of the main lower bounding methods breaks down near 
the square root of the instance size:

\begin{lemma}\label{comptriv} If the 1-certificate size of a Boolean function 
on $n$ Boolean variables is $k$
then the general weighted version of the quantum adversary method
(also known as the method of Ambainis) 
in the sense of \cite{bss} can prove only a lower bound of 
$\left( 1-2\sqrt{\epsilon(1-\epsilon)} \right)  \sqrt{nk}$.
\end{lemma}

Problems with small certificate complexity include various 
collision type problems such as the  2-1 collision problem
and the element distinctness problem.
Using the degree bound of R. Beals, H. Buhrman, R. Cleve, 
M. Mosca and Ronald de Wolf \cite{bbcm}
for the 2-1 collision problem
the first polynomial lower 
bound was shown by Aaronson \cite{aa}. Shi \cite{s}
showed tight $\Omega(n^{1/3})$ lower bound and 
Kutin \cite{k} removed the "small range assumption".
For element distinctness, an $\Omega(n^{1/2})$ 
lower bound follows by an easy
reduction from Grover's search. Shi got
 $\Omega(n^{2/3})$ and Ambainis \cite{a2} removed the 
"small range assumption." 
With a recent 
ingenious algorithm of Ambainis \cite{a}
the complexity of the element distinctness problem
was finally determined to be in $\theta(n^{2/3})$.
In a sequel of this paper 
using Ambainis's new technique
with Santha and Magniez \cite{mss}
we  could improve the present upper bound for the 
TDP to $n^{1.3}$.
The algorithm presented here is based on three combinatorial observations:

The first observation is due to H. Buhrman, C. D\"urr, M. Heiligman, 
P. H$\not${o}yer, F. Magniez, M. Santha and R. de Wolf \cite{bdhh}:

\begin{lemma}\label{many}
If graph $G_{1}$ is known then detecting a triangle with at least one edge 
$G_{1}$ can be determined with $O(\sqrt{n|G\cap G'|})$ queries.
\end{lemma}

\noindent{\bf Proof:}
The proof of the above statement 
is exactly the same as that of Theorem 8 in \cite{bdhh},
and is based on the Amplitude Amplification technique
of G. Brassard, P. Hoyer, M. Mosca, A. Tapp \cite{bhmt}. $\Box$

Note that the lemma can also be proven using Grover search,
and we loose only a factor of $\log n$
in the analysis. Throughout the paper we
do not try to optimize $\log n$ factors, since we conjecture that the
exponent we present here is not tight.
Perhaps the most crucial observation to the algorithm is the following
trivial-looking one:

\begin{lemma}\label{trivi}
For every $v\in [n]$, using
$O(n\log n)$ queries, we either find a triangle in $G$ or verify that 
$G\subseteq [n]^{2}\setminus \nu_{G}(v)^{2}$ with probability $1 - 
{1\over n^{3}}$.
\end{lemma}

\noindent{\bf Proof:} We query all edges incident to $v$ classically
using $n-1$ queries. This determines $\nu_{G}(v)$.
With Grover's search \cite{g} we find an edge of $G$ in
$\nu_{G}(v)^{2}$, if there is any. 
In order to achieve $1 - {1\over n^{3}}$ success probability
we use the following safe variant of the search:

\begin{definition}[Safe Grover Search]\label{safe}
Let $c\ge 1$ be an arbitrary constant.
The Safe Grover Search on a database of $N$ items 
is the usual Grover search iterated $O(\log N)$ times independently.
The safe Grover search uses $O_{c}(N \log N)$ queries and it 
finds the (type of) item we are looking for, if it is the database,
with probability at least $1-{1\over n^{c}}$.
\end{definition} $\Box$

This lemma with the observation that hard instances 
have to be dense, already enable us to show that the quantum 
query complexity of the TDP is $o(n^{1.5})$, using 
the lemma of Szemeredi. However another 
fairly simple observation can help us to decrease the exponent:

\begin{lemma}\label{almosttrivi}
Let $k = \lceil 4n^{\epsilon}\log n\rceil$, and let
$v_{1},v_{2},\ldots,v_{k}$
randomly chosen from $[n]$ (with no repetitions).
Let
$G' = [n]^{2}\setminus \cup_{i=1}^{k} \nu_{G}(v)^{2}$.
Then
\[
Prob_{v_{1},v_{2},\ldots,v_{k}}\left( G'\subseteq G^{\langle 
n^{1-\epsilon}\rangle }   \right) > 1- {1\over n^{3}}.
\]

\end{lemma}

\noindent{\bf Proof:} Let us first remind the reader 
about the following lemma that is useful in many applications:

\begin{lemma}\label{useful}
Let $X$ be a fixed subset of $[n]$ of size $pn$ and $Y$ be a random subset
of $[n]$ of size $qn$, where $p+q<1$. Then the probability that 
$X\cap Y$ is empty is $(1- pq)^{n(1\pm O(p^{3}+q^{3}+1/n))}$.
\end{lemma}

\noindent{\bf Proof:} The probability we are looking for 
is estimated using the Stirling formula as

\begin{eqnarray*}
{{n(1-p)\choose nq}\over {n\choose nq} } = 
{[n(1-p)]! [nq]! [n(1-q)]! \over [nq]! [n(1-p-q)]!n!} & = & \\
\sqrt{ (1-p)(1-q)\over (1-p-q)}
\left[
{(1-p)^{1-p} (1-q)^{1-q}\over(1-p-q)^{1-p-q}}\right]^{n}(1\pm o(1))
& = & (1- pq)^{n(1 \pm O(p^{3}+q^{3}+1/n))}.
\end{eqnarray*} 
$\Box$

Consider now a fixed edge $(a,b)$ such that
$t(G,a,b) \ge  n^{1-\epsilon}$. The probability that 
$(a,b)\in G'$ is the same as the probability that the set 
$X = \{x\in [n]\mid\; (x,a)\in G \,\land\,(x,b)\in G\}$
is disjoint from the 
random set $\{v_{1},v_{2},\ldots,v_{k}\}$. Notice that 
$|X|= t(G,a,b)$. By Lemma \ref{useful} we can estimate now this probability
as
\[
\left(1 - {4n^{\epsilon}\log n\over n}\times
{n^{1 - \epsilon}\over n}\right)^{
n(1+o(1))}
= \left( 1 - {4\log n\over n}\right)^{n(1+o(1))}
< e^{-3\log n} = n^{-3}.
\]
Then the lemma follows from the union bound, since the 
number of possible edges $(a,b)$ is at most $n^{2}$.$\Box$

\vspace{0.25in}

Our algorithm will have three parameters 
$\epsilon={3\over 7}$, $\epsilon'=\delta={1\over 7}$, and
it either returns a triangle during its run or it:

\smallskip

\begin{enumerate}
\item Randomly creates a graph $G'$ such that 
$G\subseteq G' \subseteq  G^{\langle 
n^{1-\epsilon}\rangle }$ with high probability;
\item In a number of steps
classifies the edges of $G'$ into $T$ and $E$
such that $T$ contains only a small number of triangles
and $E\cap G$ is small;
\item Searches for a triangle in $G$ among all triangles inside $T$;
\item Searches for a triangle of $G$ intersecting with 
$E$;
\item In case steps 3 and 4 run without
success the algorithm returns ``No.''
\end{enumerate}

\vspace{0.15in}

See the algorithm in details in Figure 1.

\begin{figure}\label{algfigure}
\begin{tabular}{l p{4.5in}c}

 & {\em Command}     &  {\em Complexity} 

\vspace{0.15in}
\\\hline\hline

1. & Set $k= \lceil 4 n^{\epsilon}\log n\rceil$
Select $\{v_{1},v_{2},\ldots,v_{k}\}$ randomly from
$[n]$ and query all edges of the form $(v_{i},v)$ for
$1\le i\le k$ and $v\in [n]$. & 
$O( n^{1+\epsilon}\log n)$ \\{}

\vspace{0.15in}

2. & For each $v_{i}$ check with the Safe Grover search 
if $G\cap \nu_{G}(v_{i})^{2}$ is empty. If some 
$1\le i\le k$
it is not, after verification 
return $(v,w), (v_{i},v), (v_{i},w)$, where
all three edges are in $G$.
 
Otherwise define 
$
G' = [n]^{2}\setminus   \cup_{i=1}^{k} \nu_{G}(v_{i})^{2}.
$
& $O( n^{1+\epsilon}(\log n)^{2})$ \\{}

\vspace{0.15in}

3. & Set $T=\emptyset$, $E=\emptyset$. & 0 \\{}

\vspace{0.15in}

4. & Find an edge $(v,w)\in G'$ for which $t(G',v,w) < 
\lceil n^{1-\epsilon'}\rceil$ if exists, and put this edge
to $T$, delete it from $G'$. Keep repeating this
until no such edge is found. & 0 \\{}

\vspace{0.15in}

5. & Pick a vertex $v$ of $G'$ with non-zero degree and 
choose one of the
{\em 1. low degree hypothesis}: $|\nu_{G}(v)| \le 
10 \lceil n^{1-\delta}\rceil$;
{\em 2. high degree hypothesis}: $|\nu_{G}(v)| \ge 
0.1 \lceil n^{1-\delta}\rceil$;
such that the chosen hypothesis holds with probability at least
$1-{1\over n^{3}}$.
& 
$O( n^{\delta}\log n)$
 \\{}

\vspace{0.15in}

6. & If in 5. we chose the low degree hypothesis, add all 
edges of $G'$ with end point $v$ into $E$
and delete them from $G'$. & 0  \\{}

\vspace{0.15in}

7. & If in 5. we chose the high degree hypothesis, 
then find all vertices in  $A \stackrel{def}{=} \nu_{G}(v)$.
First run the Safe 
Grover to
determine if there is an edge in
$\nu_{G}(v)^{2}$. If one is found, after verification, output
the triangle of $G$ induced by $v$ and this edge.
If no triangle is found add all edges
in $G'(A,A')$ to $E$, where $A'= \nu_{G'}(v)$.
Also, delete these edges from $G'$.  & $O(n\log n)$ \\{}

\vspace{0.15in}

8. & Repeat 4.-7. until $G'$ becomes empty.  & {\em See analysis} \\{}

\vspace{0.15in}

9. & Try to find a triangle inside $T$ using Grover search, verify
and output if found. If not, go to 10.  & 
$\sqrt{n^{3-\epsilon'}}\log n$ \\{}

10. & Search for a triangle in $G$ with non-empty
intersection with $E$, verify
and output if found. & 
$\sqrt{n|G\cap E|}$ \\\hline

\end{tabular}
\caption{Quantum query algorithm for the Triangle Detection Problem}
\end{figure}

\section{Analysis}

\begin{lemma}
The algorithm for the TDP described in the previous section
runs using 
\[
O(
n^{1+\epsilon}(\log n)^{2} + 
n^{1+ \delta + \epsilon'}\log n  + 
\sqrt{n^{3-\epsilon'}}\log n +
\sqrt{n^{3-\min(\delta,\epsilon-\delta-\epsilon')}}).
\]
quantum queries
and returns ``No'' with
probability one if there is no triangle in $G$, otherwise returns a
triangle of $G$ with some constant positive probability.
\end{lemma}

\noindent{\bf Proof:}

We prove the correctness and the bound on the running time
together. Clearly, if there is no triangle in the graph,
the algorithm outputs ``No,'' since the algorithm places a triplet into
the output register only after checking that it is a triangle in $G$.
Therefore the correctness proof requires only to calculate 
the probability with which the algorithm outputs a triangle
if there is any. 

\medskip

\noindent{\em Step 1:} 

\smallskip

For each $v_{i}$ we use $n-1$ deterministic queries to find
$\nu_{G}(v_{i})$. Since $1\in [k]$, where 
$k = O(n^{\epsilon}\log n)$, Step 1 costs $O(n^{1 + \epsilon}\log n)$
queries.

\medskip

\noindent{\em Step 2:} 

\smallskip

For each $v_{i}$ we use Safe Grover to find out if 
$\nu_{G}(v_{i})^{2}$ is empty. A single run of Safe Grover
takes $O(n\log n)$ queries, thus the total number of queries made is 
$O(n^{1 + \epsilon}(\log n)^{2})$.
We now claim that with probability $1-O({1\over n^{2}})$ we have that
$G\subseteq G' \subseteq  G^{\langle 
n^{1-\epsilon}\rangle }$, or else a triangle of $G$ is found.

Indeed, if no triangle is found then the Safe Grover for each $v_{i}$ 
with probability $1-{1\over n^{3}}$ verifies that
$G\cap \nu_{G}(v_{i})^{2} = \emptyset$. So $G'\supseteq G$ holds.
That $G' \subseteq  G^{\langle 
n^{1-\epsilon}\rangle }$ holds follows from Lemma \ref{almosttrivi}.

\medskip

\noindent{\em Step 4:} 

\smallskip

Realize that after steps 1. and 2.
we know $G'$, so Step 4 costs us no queries.
One of the important, but simple observations is that even
repeated applications of step 4 creates a graph $T$ with 
a small number of triangles. This is shown in the following lemma:

\begin{lemma}\label{trianglelemma}
Let $H$ be a graph on $[n]$. Assume that a graph $T$ is built
by a process that starts with an empty set, and
at every step either discards some edges from
$H$ or adds an edge $(a,b)$ of $H$ to $T$ for which
$t(H,a,b)\le\tau$ holds. For the $T$ created by the end of the process
we have $t(T)\le { n \choose 2}\tau$.
\end{lemma}

\noindent{\bf Proof:} Let us denote by $T[i]$ the edge of $T$
that $T$ acquired when it was incremented for 
the $i^{\rm th}$ time, and let us use the notation $H^{i}$ 
for the current version of $H$ before the very 
moment when $T[i] = (a_{i},b_{i})$
was copied into $T$. Since 
$\{T[i],T[i+1]\ldots \} \stackrel{def}{=} T^{i} \subseteq H^{i}$,
we have 
\[
t(T^{i},a_{i},b_{i}) \;
\le \;t(H^{i},a_{i},b_{i}) \;\le\; \tau.
\]
Now the lemma follows from
\[
t(T) = \sum_{i} t(T^{i},a_{i},b_{i}) \le 
{ n \choose 2}\tau,
\]
since $i$ can go up to at most ${ n \choose 2}$. $\Box$

\medskip

\noindent{\em Step 5:}

\smallskip

In this step we use an obvious sampling strategy:
Set a counter $C$ to 0.
Query $\lceil n^{\delta}\rceil$ random edge candidates 
from $v\times [n]$ (the constants are important).
If there is an edge of $G$ among them, 
add one to $C$. Repeat this process $K= c_{0}\log n$ times,
where $c_{0}$ is a sufficiently large constant.
Accept the low degree hypothesis if by the end $C<K/2$,
otherwise accept the large degree hypothesis. Then

\begin{enumerate}
\item The probability that $\deg_{G}(v) > 10n^{1-\delta}$
and the low degree hypothesis is accepted is at most $1/n^{3}$.
\item The probability that $\deg_{G}(v) < 0.1n^{1-\delta}$
and the high degree hypothesis is accepted is at most $1/n^{3}$.
\end{enumerate}

Indeed, using Lemma \ref{useful}, considering a single 
round of sampling the probability that our sample set 
does not contain an edge from $G$ even though
$\deg_{G}(v) > 10n^{1-\delta}$ is
\[
\left(1 - {10n^{1-\delta}\over n}\times
{n^{\delta}\over n}\right)^{
n(1+o(1))} = \left(1 - {10\over n}\right)^{
n(1+o(1))} < 0.1
\]
Similarly, the probability that our sample set 
contains an edge from $G$ even though
$\deg_{G}(v) < 0.1 n^{1-\delta}$ is 
\[
1- \left(1 - {0.1n^{1-\delta}\over n}\times
{n^{\delta}\over n}\right)^{
n(1+o(1))} = 1- \left(1 - {1\over 10n}\right)^{
n(1+o(1))} < 0.2.
\]

Now for $K= c_{0}\log n$ rounds, where $c_{0}$ is large enough
the Chernoff bound gives the claim.

\medskip

\noindent{\em Step 6:}

\smallskip

Since we know $G'$ the operation costs us no queries.

\medskip

\noindent{\em Step 7:}

\smallskip

Finding out $\nu_{G}(v)$ costs us $n-1$
classical queries. Finding out if $\nu_{G}(v)^{2}\cap G = \emptyset$
costs us $O(n\log n)$ queries, using the safe Grover.

\medskip

\noindent{\em Step 8:}

\smallskip

The key to estimating the complexity of this step
is an upper estimate on the number of executions of Step 7.
In turn, this is done by lower bounding
$|G'(A,A')|$. For each $x\in A$
we have $t(G',v,x) \ge n^{\epsilon'}$, otherwise
in Step 4. we would have classified $(v,x)$ into $T$.
A triangle $(v,x,y)$ contributing to 
$t(G',v,x)$ contributes with the edge $(x,y)$ to $G'(A,A')$.
Two triangles $(v,x,y)$and $(v,x',y')$
can give the same edge in $G'(A,A')$ only if $x=y'$ and $y=x'$.
Thus:
\begin{equation}\label{trieq}
|G'(A,A')| \ge {1\over 2} \sum_{x\in\nu_{G}(v)} t(G',v,x)
\ge   |A| n^{1-\epsilon'}/2.
\end{equation}

Since we executed Step 7 only under the large degree
hypothesis on $v$, if the hypothesis is correct,
the right hand side of Equation \ref{trieq} is at least
$0.1 n^{1-\delta}n^{1-\epsilon'}/2 = \Omega(n^{2-\delta - \epsilon'})$.
Since $G'$ has at most ${n\choose 2}$ edges, 
it can execute Step 7 at most $O(n^{\delta + \epsilon'})$ times.

How about the number of executions of Step 5? 
Every execution of Step 5 leads to either 
the execution of Step 6 or that of Step 7. 
We have already seen that the number of executions of 
Step 7 is in $n^{\delta + \epsilon'}$, which is
easily in $O(n)$ (and it would not help to choose the
parameters otherwise). We claim that 
each vertex is processed in Step 6 at most once.
Indeed, if a vertex $v$ gets into Step 6, its incident edges
are all removed, and its degree in $G'$ becomes 0
making it ineligible for being processed in Step 5 again.
Thus the total complexity of Step 8 is:
\[
n^{\delta + \epsilon'} O(n\log n) + O(n) O(n^{\delta}\log n).
\]

\medskip

\noindent{\em Step 9:}

\smallskip

By Lemma \ref{trianglelemma} we have that there are at most 
$n^{2}n^{1-\epsilon'}$ triangles are in $T$. 
$T$ is a graph that is known to us, and so we can
find out if one of these triangles belong to $G$ 
in time $O(\sqrt{n^{3-\epsilon'}}\log n)$, using Safe Grover.

\medskip

\noindent{\em Step 10:}

\smallskip

In order to estimate $G\cap E$ observe that we
added edges to $E$ only in Steps 6. and 7.
Assume the Safe Grover worked throughout the whole algorithm.
In each execution of Step 6. we added at most 
$10 n^{1-\delta}$ edges to $E$, and we had $O(n)$
such executions that give a total of 
$O(n^{2-\delta})$ edges. The number of executions of Step 7
is $O(n^{\delta + \epsilon'})$. Our task is now to bound
now the the number of edges of $G$ each such execution
adds to $E$.

We estimate $|G\cap G'(A,A')|$ from the $A'$ side.
This is the only place where we use the fact that 
$G' \subseteq  G^{\langle 
n^{1-\epsilon}\rangle }$. For every $x\in A'$ we have 
$t(G,v,x)\le n^{1-\epsilon}$.
On the other hand every edge $(y,x)$ $y\in A$, $x\in A'$, $(y,x)\in G'$
creates a $(v,x)$-based triangle.
Thus 
\[
|G\cap G'(A,A')| \le |A'| n^{1-\epsilon} \le n^{2-\epsilon}.
\]

Therefore the total number of edges of $G$ Step 7. contributes to $E$
is $n^{2-\epsilon+\delta+\epsilon'}$.
In conclusion,
\[
|G\cap E|\le O(n^{2-\delta}+n^{2-\epsilon+\delta+\epsilon'}).
\]

By Lemma \ref{many} the complexity of finding a triangle in $G$
that contains an edge from $E$ is 
$O\left(\sqrt{n^{3-\min(\delta,\epsilon
-\delta-\epsilon')}}\right)$.

\subsection{Conclusions of the Analysis}

The probability that not all the Safe Grovers run correctly can be 
upper bounded by $1\over n$, for large $n$, using the union bound.
Thus in the sequel we assume that none of the Safe Grovers fails.
{}From the analysis we conclude that the total 
number of queries is upper bounded by:

\[
O( n^{1+\epsilon}\log n + 
n^{1+\epsilon}(\log n)^{2} + 
(n^{1+ \delta + \epsilon'}\log n + n^{1+\delta}\log n) + 
\sqrt{n^{3-\epsilon'}}\log n +
\sqrt{n^{3-\min(\delta,\epsilon-\delta-\epsilon')}}).
\]

With $\epsilon={3\over 7}$, $\epsilon'=\delta={1\over 7}$
this gives $O(n^{1+{3\over 7}}\log^{2}n)$ for the total number of queries.
When the Safe Grover does not fail, 
the structural assumptions we made 
based on it hold.
$G'$ eventually lends all its edges 
to $T$ and $E$. Since $G\subseteq G'$,
every triangle in $G$ either has to be contained totally in $T$
or it has to have a non-empty intersection with $E$.
Finally, we mention that in case one of the Safe Grovers goes wrong, 
the algorithm cannot run beyond the given time bound:
quantum query machines by definition
execute a prescribed number of steps.

\medskip

\noindent{\em Acknowledgments:} The author thanks 
Fr\'ed\'eric Magniez and Miklos Santha for bringing the problem
and Lemma \ref{many} to his attention and Andris Ambainis for references.

\newpage

\section{Appendix}

\noindent{\bf Proof of Lemma \ref{comptriv}:}\\

\medskip

For a matrix $M$, let $\lambda(M)$ denote the greatest
eigenvalue of $M$. Recall the general form of Ambainis from
\cite{bss}:

\begin{theorem}
\label{thm:ev lb}
Let $f:\inD \longrightarrow \outD$ be a partial boolean function
with $\inD \subseteq \{0,1\}^n$.  Let $\Gamma$ be an arbitrary
$\inD \times \inD$ nonnegative symmetric matrix that satisfies $\Gamma[x,y]=0$
whenever $f(x)=f(y)$. For $i \in \{1,\ldots,n\}$ let
$\Gamma_i$ be the matrix:

\[
\Gamma_i[x,y]=\left\{
\begin{array}{l}
 \mbox{0 if $x_i=y_i$}\\
\mbox{$\Gamma[x,y]$ if $x_i \neq y_i$}.
\end{array}
\right.
\]
Then:
\[
\DQ_{\epsilon}(f) \geq  
{ \left(1-2\sqrt{\epsilon(1-\epsilon)}\right)\lambda(\Gamma) \over
2\max_{1 \leq i \leq n} \lambda(\Gamma_i) }.
\]
\end{theorem}

Above $\DQ_{\epsilon}(f)$ denotes the $\epsilon$ error quantum 
query complexity of $f$. We need to work in the case when 
$\inD\subseteq\Sigma^{n}$ and
$\outD = \{0,1\}$.
Let $A = f^{-1}(1)$.
Let $k$ be the 1-certificate size of $f$.
For an input $x\in A$ let $A_{x}$ be its 
1-certificate. Let

\[
\Gamma[x,y]'=\left\{
\begin{array}{l}
 \mbox{$\Gamma[x,y]$ if $f(x)= 1$ and $f(y)= 0$}\\
\mbox{0 otherwise}.
\end{array}
\right.
\]

Let 
$v\in {\bf R}^{D}$ be the unit vector with 
$\lambda(\Gamma) = v\Gamma v^{\ast}$. Notice that $v$ is positive.
Because $\Gamma[x,y]$ is 0
whenever $f(x) = f(y)$ and $\Gamma$ is symmetrical, it holds that
$v\Gamma' v^{\ast} = \lambda/2$.

For 
$1\le i\le n$ let $v_{i}$ be the vector which equals to 
$v$ on those coordinates $x$ where $i\in A_{x}$ and 0 otherwise.
We now have:

\begin{eqnarray}\label{firsteq}
\langle v_{i}, v\rangle & = & \langle v_{i}, v_{i}\rangle\;\;\; \mbox{
for $1\le i\le n$.} \\\label{eqtwo}
v_{1}+\ldots +v_{n} & \le & kv; \\\label{gammaeq}
v_{1}\Gamma_{1}v + v_{2}\Gamma_{2}v + \ldots + v_{n}\Gamma_{n}v & \ge & 
v\Gamma'v = \lambda/2; \\\label{lasteq}
{\lambda(\Gamma)\over \max_{1 \leq i \leq n} \lambda(\Gamma_i)}
& \le & 2 (|v_{1}|+\ldots + |v_{n}|).
\end{eqnarray}

Here the first inequality is meant entry-wise, and Inequality
(\ref{lasteq}) follows from Inequality (\ref{gammaeq})
and from $v_{i}\Gamma_{i}v = |v_{i}|{v_{i}\over |v_{i}|} \Gamma_{i}v \le 
|v_{i}|\lambda(\Gamma_{i})$.
Now from (\ref{firsteq}) and (\ref{eqtwo}) it follows that
\[
\sum_{i=1}^{n} |v_{i}| \;\le\; \sqrt{n \sum_{i=1}^{n} |v_{i}|^{2}}
\; =\;  \sqrt{n \sum_{i=1}^{n} \langle v_{i} | v\rangle}\;
\le\; \sqrt{n \langle kv|v\rangle } = \sqrt{n k}.
\]
The above together with (\ref{lasteq}) gives the lemma. $\Box$


\begin{thebibliography}{BDHH}

\bibitem[Aa]{aa}
Scott Aaronson: Quantum lower bound for 
the collision problem. STOC 2002: 635-642

\bibitem[A]{a2} 
Ambainis: 
Quantum Lower Bounds for Collision and Element Distinctness with Small Range.
quant-ph/0305179.  

\bibitem[A2]{a} Ambainis: lecture on the complexity of collision finding
(communicated by Fr\'ed\'eric Magniez)

\bibitem[BSS]{bss}
Howard Barnum and Michael Saks and Mario Szegedy:
Quantum decision trees and semidefinite programming.
CCCC 2003

\bibitem[BBCM]{bbcm}
Robert Beals, Harry Buhrman, Richard Cleve, 
Michele Mosca, Ronald de Wolf: Quantum Lower 
Bounds by Polynomials. FOCS 1998: 352-361

\bibitem[BCWZ]{bcwz}
Harry Buhrman, Richard Cleve, Ronald de Wolf, 
Christof Zalka: Bounds for Small-Error and 
Zero-Error Quantum Algorithms. FOCS 1999: 358-368

\bibitem[BHMT]{bhmt}
Gilles Brassard, Peter Hoyer, Michele Mosca, Alain Tapp:
Quantum Amplitude Amplification and Estimation.
quant-ph/0005055 

\bibitem[BDHH]{bdhh}
Harry Buhrman, Christoph D\"urr, Mark Heiligman, Peter Hoyer, 
Fr\'ed\'eric Magniez, Miklos Santha, Ronald de Wolf: Quantum Algorithms for
Element Distinctness. IEEE Conference on Computational 
Complexity 2001: 131-137

\bibitem[CK]{ck}
Amit Chakrabarti, Subhash Khot: Improved Lower 
Bounds on the Randomized Complexity of Graph Properties. 
ICALP 2001: 285-296

\bibitem[G]{g}
Lov K. Grover: A Fast Quantum Mechanical Algorithm for 
Database Search. STOC 1996: 212-219

\bibitem[H]{h}
Peter Hajnal:
An $n^{4/3}$ lower bound on the randomized 
complexity of graph properties, Combinatorica, 11(1991), 131--143

\bibitem[LY]{ly}
L. Lov\'asz and N. Young. Lecture notes on evasiveness 
of graph properties. Technical Report CS-TR-317-91, 
Computer Science Department, Princeton
 University, 1991.


\bibitem[K]{k}
Samuel Kutin
A quantum lower bound for the collision problem
quant-ph/0304162

\bibitem[MSS]{mss}
F. Magniez, M. Santha, and M. Szegedy: An O(n1.3) 
quantum algorithm for the triangle problem. 
Technical Report quant-ph/0310134, arXiv, 2003.

\bibitem[S]{s}
Yaoyun Shi: Quantum Lower Bounds for the 
Collision and the Element Distinctness Problems. FOCS 2002: 513-519

\end{thebibliography}
\end{document}